\documentclass{aastex61}
\newcommand\aastex{AAS\TeX}

\accepted{July 5, 2018}
\submitjournal{ApJL}

\shorttitle{\aastex\ Solar cycle 25 detected in zonal flows}
\shortauthors{Howe et al.}

\begin{document}

\title{Signatures of solar cycle 25 in subsurface zonal flows}

\correspondingauthor{R. Howe}
\email{r.howe@bham.ac.uk}

\author[0000-0002-3834-8585]{R. Howe}
\affil{School of Physics and Astronomy, 
University of Birmingham, 
Edgbaston, Birmingham B15 2TT, UK}

\affil{Stellar Astrophysics Centre (SAC), Department of Physics and Astronomy, Aarhus University, Ny
Munkegade 120, DK-8000 Aarhus C, Denmark}

\author{F. Hill}
\affiliation{National Solar Observatory, 3665 Discovery Drive, 3rd Floor
Boulder, CO 80303, USA}

\author{R. Komm}
\affiliation{National Solar Observatory, 3665 Discovery Drive, 3rd Floor
Boulder, CO 80303, USA}

\author{W.~J. Chaplin}
\affil{School of Physics and Astronomy, 
University of Birmingham, 
Edgbaston, Birmingham B15 2TT, UK}

\affil{Stellar Astrophysics Centre (SAC), Department of Physics and Astronomy, Aarhus University, Ny
Munkegade 120, DK-8000 Aarhus C, Denmark}

\author{Y. Elsworth}
\affil{School of Physics and Astronomy,
University of Birmingham, 
Edgbaston, Birmingham B15 2TT, UK}

\affil{Stellar Astrophysics Centre (SAC), Department of Physics and Astronomy, Aarhus University, Ny
Munkegade 120, DK-8000 Aarhus C, Denmark}

\author{G.~R. Davies}
\affil{School of Physics and Astronomy,
University of Birmingham,
Edgbaston, Birmingham B15 2TT, UK}

\affil{Stellar Astrophysics Centre (SAC), Department of Physics and Astronomy, Aarhus University, Ny
Munkegade 120, DK-8000 Aarhus C, Denmark}

\author{J. Schou}
\affil{Max-Planck-Institut f{\"u}r Sonnensystemforschung,
Justus-von-Liebig-Weg 3,
37077 G{\"o}ttingen,
Germany}


\author{M.~J. Thompson}
\affil{High Altitude Observatory/National Center for Atmospheric Research,
 P.O. Box 3000, Boulder, CO 80307-3000, USA}



\begin{abstract}

The pattern of migrating zonal flow bands associated with the solar cycle, 
known as the torsional oscillation, has been monitored with continuous global helioseismic observations by the Global Oscillations Network Group, together with those made by 
the {\em Michelson Doppler Imager} onboard the {\em Solar and Heliosepheric Observatory} and its successor the {\em Helioseismic and Magnetic Imager} onboard the {\em Solar Dynamics Observatory}, since 1995, giving us nearly two full solar cycles of observations. We report that the flows now show traces of the mid-latitude acceleration that is expected to become the main equatorward-moving branch of the zonal flow pattern for Cycle 25. Based on the current position of this branch, we speculate that the onset of widespread activity for Cycle 25 is unlikely to be earlier than the middle of 2019.

\end{abstract}




\section{Introduction} \label{sec:intro}

While the solar cycle is defined by the growth, migration and decay of surface magnetic activity, it also has a more subtle manifestation in the pattern of migrating bands of faster and slower rotation known as the torsional oscillation. This pattern was first observed in surface Doppler measurements at the Mount Wilson observatory by \citet{1980ApJ...239L..33H}. With the advent of continuous monitoring of medium-degree solar acoustic modes by the Global Oscillation Network Group \citep[GONG:][]{1996Sci...272.1292H} and the {\em Michelson Doppler Imager} \citep[MDI:][]{1995SoPh..162..129S} it became possible to follow the migrating flows below the surface by subtracting the average rotation profile from a series of two-dimensional rotation maps \citep{2000ApJ...533L.163H,2000ApJ...541..442A,2002Sci...296..101V}. While the zonal flow modulation is generally believed to be a tracer rather than a driver of the solar cycle, acceleration in the flows at a given latitude can be seen in advance of the appearance of large-scale magnetic activity \citep{2017SoPh..292..122K}. This may be related to the changes in the toroidal magnetic field in advance of the new-cycle activity reported by \citet*{2010ASPC..428..109L}, and it provides a tool for anticipating the 
new cycle before sunspots appear. There is also evidence for an extended cycle at high latitudes in the corona \citep{2013SoPh..282..249T}.

\begin{figure}

\includegraphics[width=\linewidth]{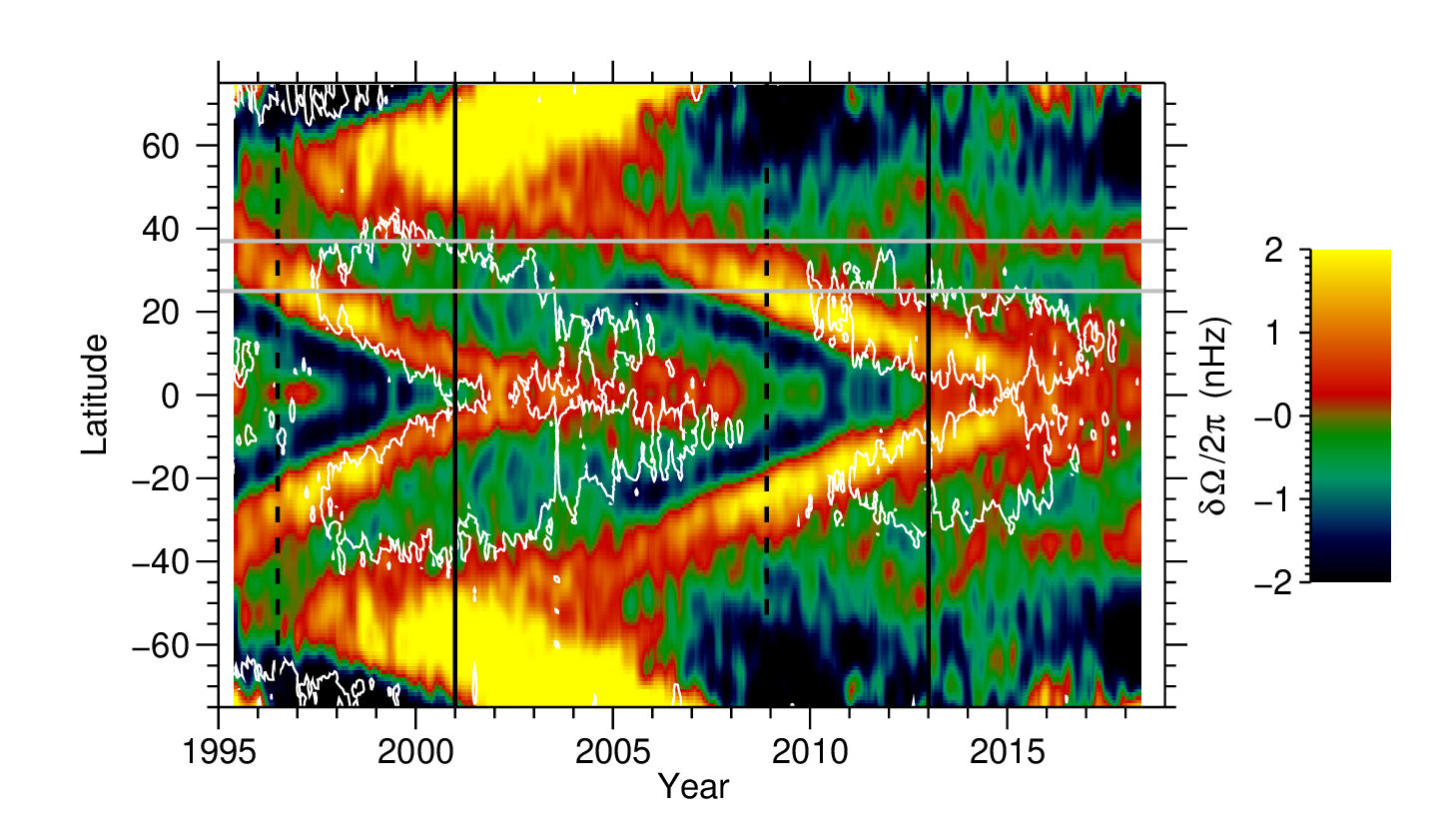}
\caption{Rotation-rate residuals at a target depth of $0.99\,R_{\mathrm{SUN}}$ as a function of latitude and time, from RLS inversions of GONG, MDI, and HMI data. The mean to be subtracted was taken separately over the whole data set for GONG and for the combined MDI and HMI set. The vertical black lines represent the times of solar minimum ({\em dashed}) and solar maximum ({\em solid}). The horizontal gray lines indicate the 25 and 37 degree latitudes. The white contours represent 10\,\% of the maximum level of the synoptic unsigned magnetic field strength.}
\label{fig:fig1}
\end{figure}

The pattern known as the torsional oscillation shows some variation from cycle to cycle, but the main features are consistent. In general, the equatorward-propagating belt of faster flow seems to appear at 
a latitude of 
about $40 \deg$ 
a  year or two after the maximum of the previous cycle and then moves towards the equator along with the new-cycle activity, finally disappearing around the time of solar minimum; see, for example, \citet{2011JPhCS.271a2074H}, where the helioseismic record is extended back in time using the Mount Wilson Doppler observations. The appearance of widespread activity in the new cycle \citep[sometimes referred to as the ``onset'' of the cycle; see, for example][]{2004AAS...204.3907S} has historically coincided with this equatorward-moving belt reaching a latitude of about $25 \deg$ \citep{2009ApJ...701L..87H,2011JPhCS.271a2074H}. There is also usually a strong poleward-moving branch that starts at about the same time as the equatorward branch; this was first pointed out by \citet{2001ApJ...559L..67A} and shows very clearly in the helioseismic data for Cycle 23 (see Figure~\ref{fig:fig1}). During the extended minimum following Cycle 23, the new equatorward branch was visible as expected, but it moved more slowly during the declining phase of Cycle 23 than the corresponding feature in the previous cycle \citep{2009ApJ...701L..87H}, resulting in an effective length of about 12.3 years for Cycle 23. During Cycle 24, the poleward branch was unusually weak, and indeed it is hardly visible in a conventional torsional-oscillation plot such as that in Figure~\ref{fig:fig1} because it is superimposed on a slower overall rotation at high latitude \citep{2013ApJ...767L..20H}, which may be related to the weaker polar fields in Cycle 24 \citep{2012ApJ...750L...8R}. 

With Cycle 24 well into its declining phase, and with nearly 23 years of continuous observations available from GONG, MDI, and the successor to MDI, the {\em Helioseismic and Magnetic Imager} \citep[HMI:][]{2012SoPh..275..229S} onboard the {\em Solar Dynamics Observatory}, we revisit the torsional oscillation data in search of the early signs of Cycle 25.

\begin{figure}

\includegraphics[width=0.5\linewidth]{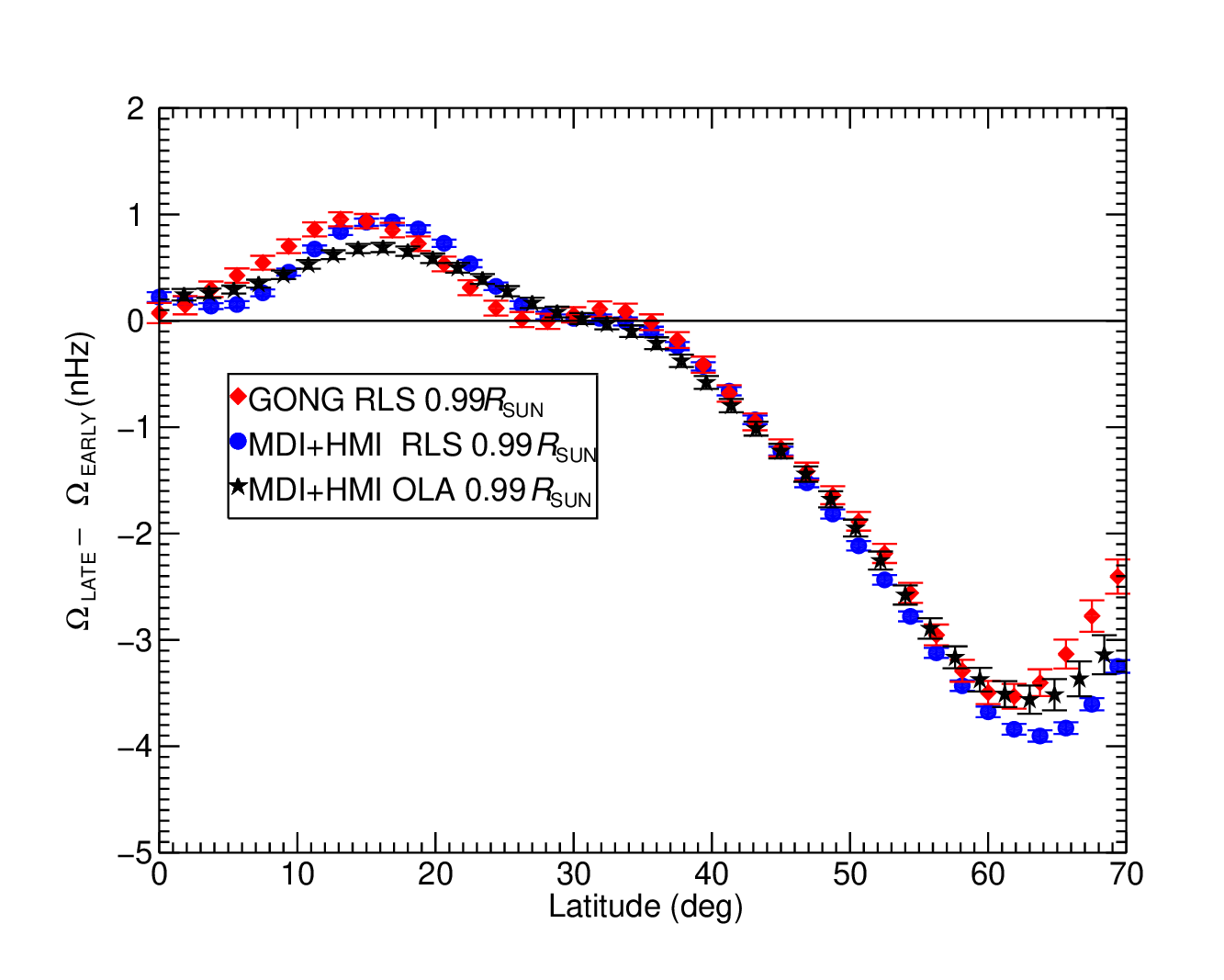}
\includegraphics[width=0.5\linewidth]{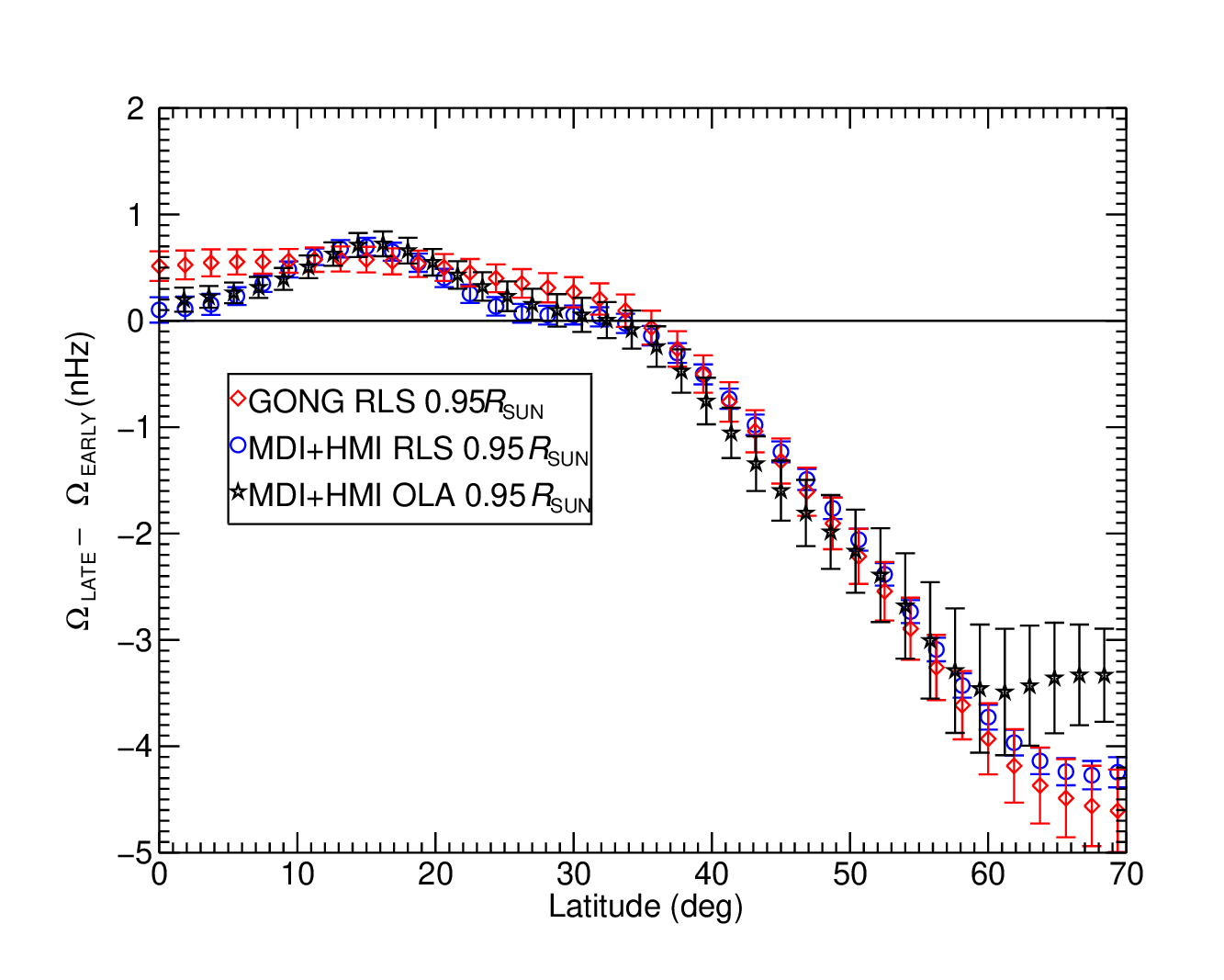}

\caption{Differences $\Omega_{\mathrm LATE}-\Omega_{\mathrm EARLY}$ between the mean inferred solar rotation rate
for 8.5-year averages starting at 1995.5 (`EARLY') and 2007.9 (`LATE'), plotted as a function of latitude at depths of ({\em left}) 0.99  and ({\em right}) 0.95 $R_{\mathrm{SUN}}$.
}
\label{fig:fig2a}
\end{figure}

\begin{figure}

\includegraphics[width=\linewidth]{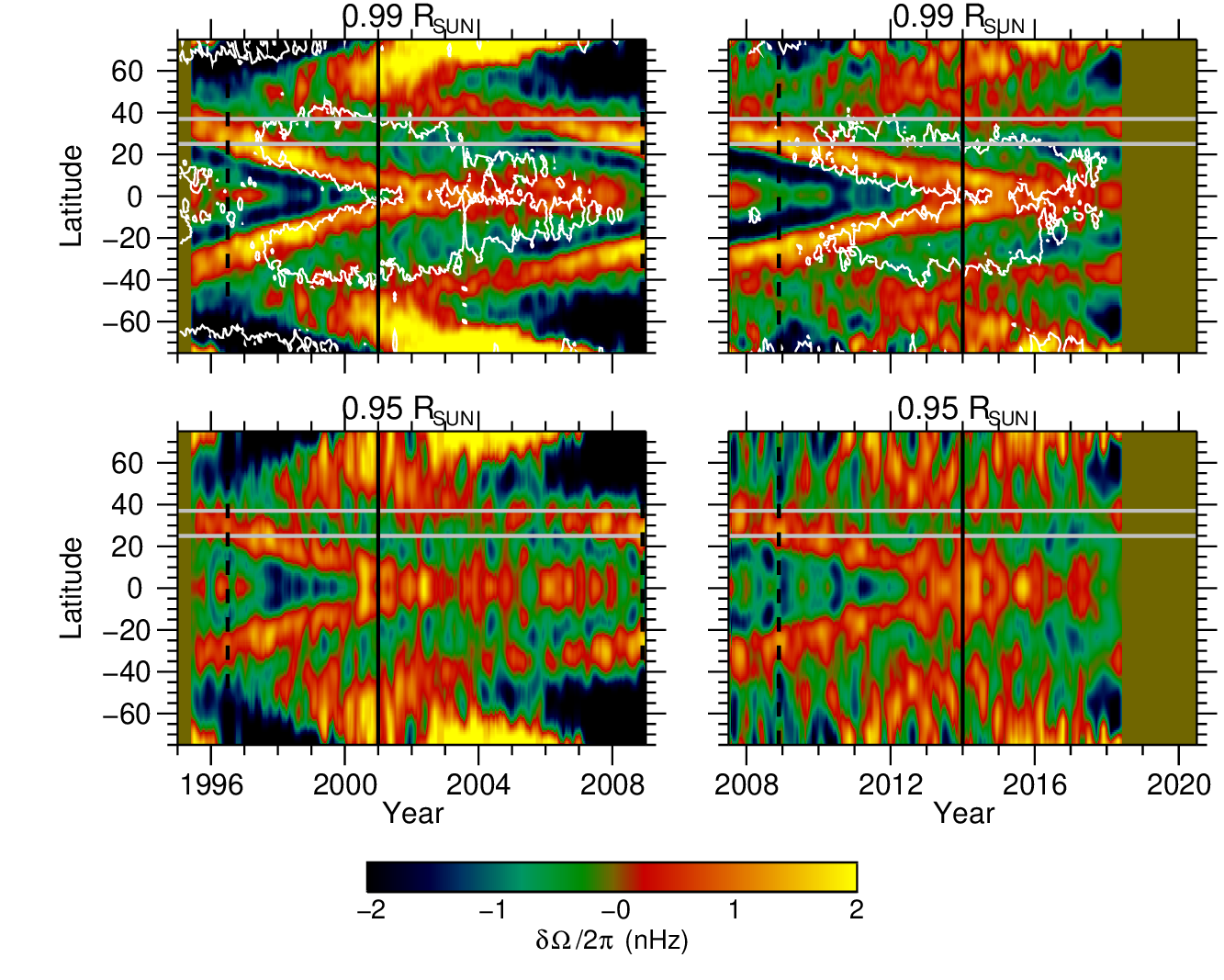}
\caption{Rotation rate residuals as a function of time and latitude for Cycle 23 (left) and Cycle 24 (right) at $0.99\,R_{\mathrm{SUN}}$ (top) and $0.95\,R_{\mathrm{SUN}}$ (bottom) with the mean taken over the 8.5 years of observations  starting a year before solar minimum, for GONG, MDI, and HMI RLS inversions. The vertical black lines represent the times of solar minimum ({\em dashed}) and solar maximum ({\em solid}). The horizontal gray lines indicate the 25 and 37 degree latitudes. The white contour on the top plots represents 10\,\% of the maximum value of the unsigned NSO magnetic field strength for the period covered by each plot.}
\label{fig:fig2}
\end{figure}

\begin{figure}
\begin{center}

\includegraphics[width=0.7\linewidth]{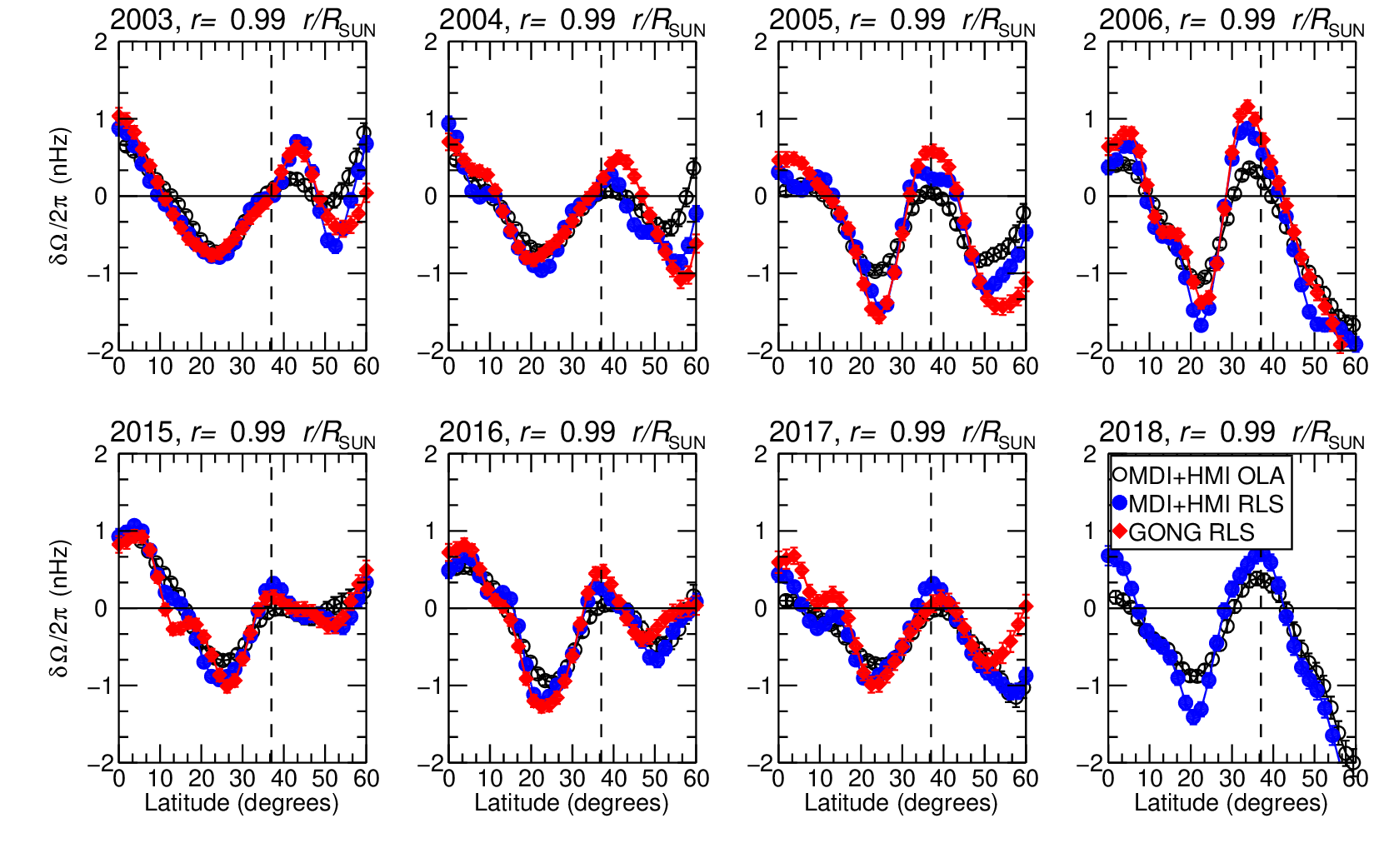}

\includegraphics[width=0.7\linewidth]{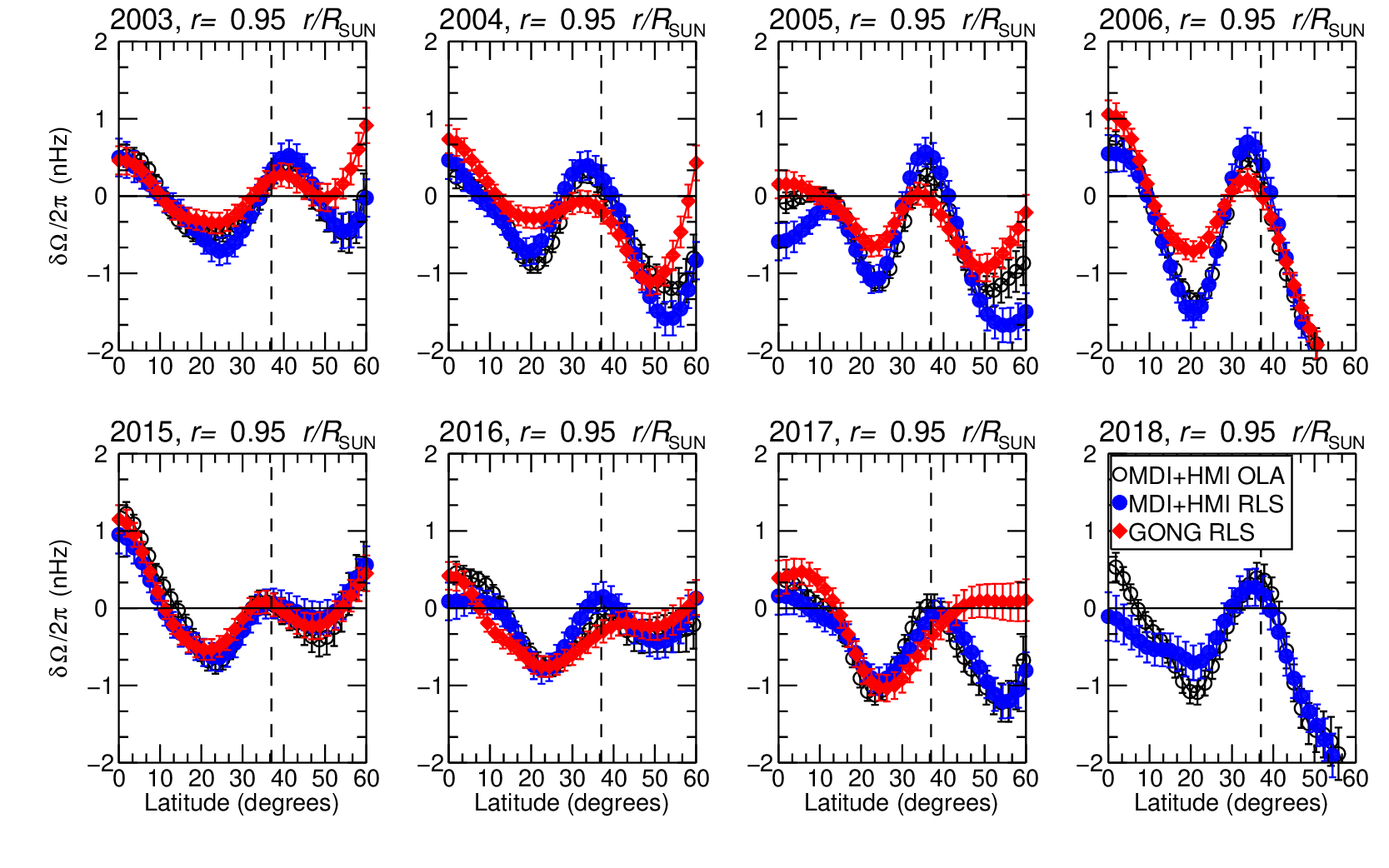}
\end{center}
\caption{Rotation-rate residuals as a function of latitude at $0.99\,R_{\mathrm{SUN}}$ (top two rows) and $0.95R_{\mathrm{SUN}}$ (bottom two rows), averaged over 1-year periods centered on 2003, 2004, 2005, 2006, and the corresponding epochs 12 years later. The vertical dashed line marks the 37 degree latitude.The residuals were calculated using a mean over the 8.5 years starting a year prior to the minimum of the corresponding solar cycle. The 2017 data for GONG include only the first half of the year: the 2018 epoch includes only the first five months of the year for HMI and no GONG data.}
\label{fig:fig3}
\end{figure}

\begin{figure}
\begin{center}

\includegraphics[width=0.4\linewidth]{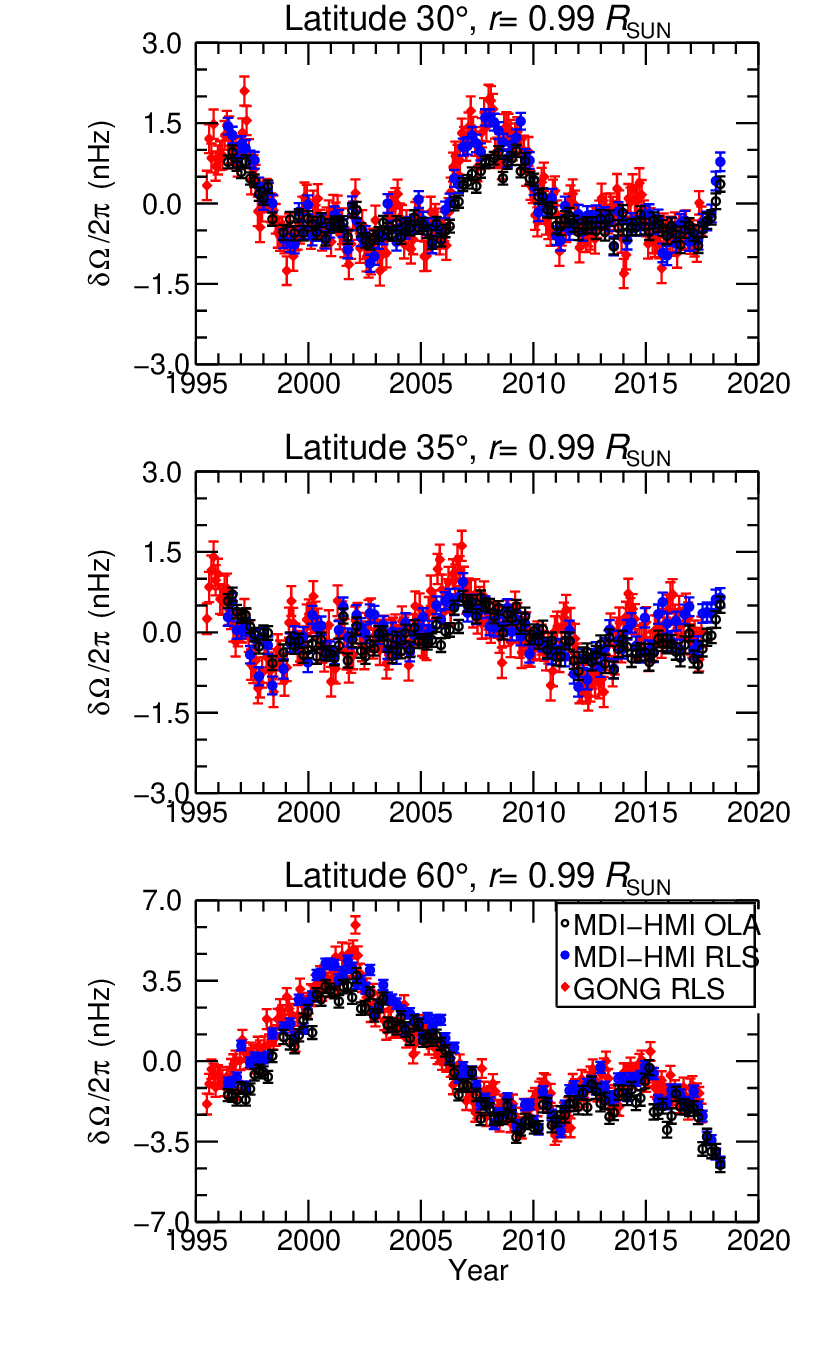}
\includegraphics[width=0.4\linewidth]{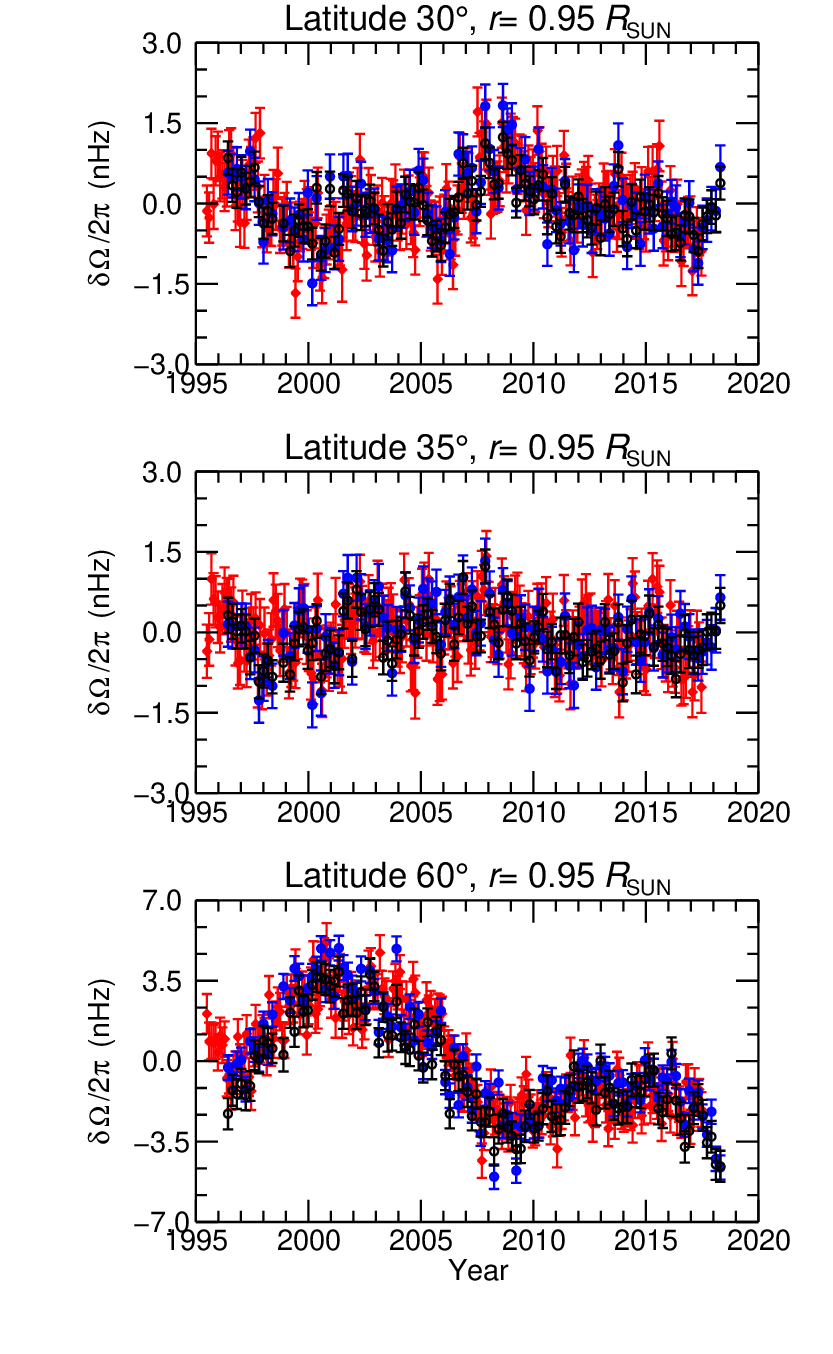}

\end{center}
\caption{Residuals of rotation rate from RLS inversions of GONG and MDI--HMI data and OLA inversions of MDI--HMI, for selected latitudes and depths.}
\label{fig:fig4}
\end{figure}

\section{Data and analysis}
As in our previous work, we analyze 2d rotation profiles inferred by two different inversion codes -- regularized least squares (RLS) and optimally localized averaging (OLA) -- based on frequencies derived from GONG data in 108-day time series with start dates at 36-day intervals together with MDI and HMI data in 72-day time series that do not overlap. We analyzed 224 GONG data sets of rotational splitting values for degrees $l \leq 150$, with the first starting in May 1995 and the last ending in August 2017. For MDI we considered 74 sets of rotational-splitting coefficients up to $l=300$, from May 1996 to February 2011, while the 41 sets of HMI observations start in  April 2010 and end with May 2018. As in \citet{2013ApJ...767L..20H}, the one-year overlap between MDI and HMI was used to remove an offset between the results of the two instruments due to systematic issues with the MDI coefficients \citep{2002ApJ...567.1234S,2009ASPC..416..311L,2015SoPh..290.3221L}.

To compare the flow patterns with the migration of magnetic activity we use 
the unsigned field strength taken from National Solar Observatory (NSO) synoptic magnetic observations \citep[for details see][]{2017MNRAS.464.4777H}.

\section{Results}

Figure~\ref{fig:fig1} shows the rotation-rate residuals from RLS inversions of GONG, MDI, and HMI data at a target depth of $0.99\,R_{\mathrm{SUN}}$. The temporal mean was taken over the whole of the GONG and MDI--HMI data sets separately before the residuals were combined. As is conventional in this type of plot, the inversion data, which do not distinguish between the northern and southern hemispheres, have been reflected in the equator, whereas the over-plotted contours of magnetic field strength have not been symmetrized. Here and throughout, the zonal flows are shown in terms of rotational frequency $\nu=\Omega/2\pi$, where $\Omega$ is the angular velocity; the corresponding linear speed at radius $r$ and co-latitude $\theta$ can be calculated using $v=r\Omega \sin{\theta}$. Close to the surface, a rotation-rate change of 1 nHz translates to a linear speed of roughly 4.4 m\, s$^{-1}$ at the equator, 3.8 m\, s$^{-1}$ at $30 \deg$ latitude, and 2.2  m\, s$^{-1}$ at $60\deg$ latitude.

The equatorward branch for Cycle 24 can be traced from around 2004\,--\,2005, where it separates from the strong poleward branch of Cycle 23, to the most recent observations in 2018, where it is close to the equator but still strong; indeed, this branch has faster flows relative to the mean than the Cycle 23 branch at the equivalent epoch, even though the magnetic activity in Cycle 23 was stronger. The poleward branch for Cycle 24 is barely visible in this representation. We do, however, see traces of a weak flow band 
at around 37 degrees latitude from 2015 onward. This band 
can perhaps be traced back as far as 2012 at higher latitudes, although this latter point is open to interpretation. In the most recent data (early 2018), this band seems to have strengthened and shifted slightly closer to the equator.

Because the underlying rotation rate in the two cycles is different, we take the same approach as \citet{2013ApJ...767L..20H} and prepare the residuals for each cycle with a tailor-made mean profile.  For the current work we have chosen to take the mean over the 8.5 years starting a year before solar minimum for each cycle, because this is the longest interval for which we have data for the equivalent epoch of both cycles. The difference in the means, as a function of latitude, is shown in Figure~\ref{fig:fig2a}. The most notable feature is the decrease in the rotation rate between 40 and 60 degrees latitude, which is common to all three of the data sets and was also seen in the similar analysis (with a shorter averaging time) of \citet{2013ApJ...767L..20H}. We also see a smaller increase for all of the data sets 
in the mean rotation rate below 20 degrees, which corresponds to the stronger low-latitude zonal flow seen in Figure~\ref{fig:fig1} in the later stages of Cycle 24. This speed-up, which was less evident in the \citet{2013ApJ...767L..20H} analysis, indicates some redistribution of angular momentum to the lower latitudes.
The differences among the data sets probably arise because of the different systematic errors from the different inversion techniques and the different uncertainties on the input data.

In Figure~\ref{fig:fig2} we compare the residuals for the two cycles computed by subtracting the 8.5-year means. We show the residuals as a function of time and latitude in two 14--year windows, with the mean subtracted over the 8.5 years starting one year before the minimum of the solar cycle. The new-cycle branch at around 37 degrees is more clearly seen when the data are plotted in this way; we also see the poleward branch for Cycle 24 more clearly, although it is still relatively weak and noisy. In addition, from 2017 onward we see a strengthening region of slower-than-average rotation poleward of the new-cycle branch, as the positive poleward branch has moved out of the latitude range where we can make reliable measurements.

To verify that the feature at 37 degrees is not just an artifact of the color table, in Figure~\ref{fig:fig3} we show the yearly-averaged residuals as a function of latitude at selected depths for GONG RLS, MDI/HMI RLS, and MDI/HMI OLA at dates centered on the midpoints of the years 2003 to 2006 and the corresponding dates 12 years later, again with the mean subtracted over the 8.5 years starting a year before the minimum of the corresponding cycle. At both epochs, we see a similar pattern with local maxima close to the equator and at around 35\,--\,40 degrees latitude and a local minimum at about 25 degrees.  At the $0.95 R_{\mathrm{SUN}}$ depth the local maximum in the recent data is less clear and does not rise above zero until 2018.  In Cycle 23 the mid-latitude branch shows clear migration at the $0.99\,R_{\mathrm{SUN}}$ depth over the four years, while in the Cycle 24 data the maximum becomes more distinct but shows little migration. 
The 2004\,--\,2006 data show a strong deceleration at higher latitudes as the poleward branch of faster rotation ends. A similar but less pronounced pattern is seen in 2015\,--\,2018, because the poleward branch is so much weaker, but we see a deepening (and poleward-moving) minimum, which by 2018 has moved beyond $60 \deg$.   
As in Figure~\ref{fig:fig2a}, the differences among the data sets at each epoch are due to their different resolution and noise properties, but the main features are clear in all. The high-latitude discrepancy between GONG and HMI in the 2017 panel is due to the rapid changes that took place in the second half of the year, which is covered only by the HMI data. Figure~\ref{fig:fig4} shows the rotation-rate residuals as a function of time at depths of  $0.99$ and $0.95\,R_{\mathrm{SUN}}$ at latitudes of 30, 35, and 60\,$\deg$ to illustrate these recent changes. These residuals were calculated by subtracting a mean over the whole data set at each latitude and depth. It is interesting to note that the new deceleration at $60 \deg$ brings the near-surface rotation rate there to its lowest level in our 23 years of observations.

From  Figures~\ref{fig:fig2} and \ref{fig:fig3} we can see that the position of the presumed Cycle-25 equatorward branch in 2016\,--\,2017 corresponds to that of Cycle 23 in 1995 and Cycle 24 in 2005 (that is, about 1.5 years and 3.5 years before solar minimum, respectively). The sharp acceleration seen at $30 \deg$ latitude in Figure~\ref{fig:fig4} starts in mid-2017 and seems to correspond to the acceleration at the same latitude starting in early 2006.%

\section{Discussion and conclusions}

We have analysed nearly 23 years of helioseismic observations of migrating zonal flow bands in the solar convection zone. The finding of \citet{2013ApJ...767L..20H} that the rotation rate at higher latitudes has been slower in Cycle 24 than in Cycle 23 has been confirmed. Such a slowing effect was linked by \citet{2013ApJ...767L..20H} to weaker polar fields. As of our most recent observations there has been a further drop in the rotation rate at 60 degrees latitude, bringing it to its lowest level since the start of the GONG measurements.

We see evidence for a weak band of faster-than-average flow that remained at a latitude of around 35\,--\,37 degrees during 2015\,--\,2017; in the most recent HMI observations from the first half of 2018 this band strengthened and shifted slightly closer to the equator, resulting in a sharp acceleration at 30 degrees. At the same time there was a sharp deceleration at 60 degrees latitude. We believe these features are associated with the upcoming Cycle 25. However, even considering the different underlying profile, the
new branch is weak compared to the one that was seen at this latitude before the onset of Cycle 24, particularly at greater depth. 

The acceleration at 30 degrees from the second half of 2017 onward comes 11.5 to 12 years after the one in the previous cycle. Can we use this to estimate the timing of solar minimum and the onset of activity in the next cycle? It is tempting but dangerous to say that on this basis we could expect the new minimum 11.5 to 12 years after the previous one (that is, in mid-to-late 2020). The danger arises for two reasons. The first issue is that the new-cycle branch is associated with new-cycle activity rather than with the fading of old-cycle activity; the second is that the period between 2006 and the present includes the unusually extended solar minimum that followed Cycle 23, which manifested in the flows as a delay in the propagation of the equatorward branch between 35 and 25 degrees latitude, so that we cannot use it to project future behavior without assuming that the upcoming minimun will be similarly prolonged -- an assumption that cannot easily be justified. 

In the historic data from Mount Wilson \citep[see, for example][]{2011JPhCS.271a2074H}, in each of Cycles 21, 22, and 23 activity became widespread around the time that the new equatorward branch reached the 20\,--\,25-degree latitude range, and the average time for the branch to migrate from 35 to 25 degrees was slightly less than two years, while before the onset of Cycle 24 the same migration took about three years. If we take the two-year time as typical and assume that the Cycle 25 branch started to move from 35 degrees in mid-2017, it seems possible that it could reach 25 degrees as early as mid-2019, but probably no sooner than that. A 2019 onset date for Cycle 25 would not be consistent with a 2020 minimum; while it would be consistent with a 22-year combined length for Cycles 23 and 24, it would be less than ten years after the onset of Cycle 24 in late 2009, and that ten-year offset would in turn suggest a late-2018 minimum. The situation should become clearer with the next year or so of observations, and we will continue to monitor the flows and update in future work.  

While the high-latitude part of the torsional oscillation pattern has been unusually weak in Cycle 24, the low-latitude branch appears stronger than that seen in Cycle 23.  The Cycle 23 flows in turn were about the same strength as those seen in Cycle 22 in Mount Wilson Doppler observations \citep{2006SoPh..235....1H}, even though the Cycle 23 activity level was somewhat lower than that in Cycle 22. 
This suggests that there is not a positive correlation between the activity level and the strength of the flows, although it is difficult to claim a negative correlation based on three data points. Therefore, it is not possible to confidently predict the strength of Cycle 25 based on the currently weak signature of the flows associated with it.

\acknowledgements
This work utilizes data obtained by the Global Oscillation Network
Group (GONG) program, managed by the National Solar Observatory, which
is operated by AURA, Inc. under a cooperative agreement with the
National Science Foundation. The data were acquired by instruments
operated by the Big Bear Solar Observatory, High Altitude Observatory,
Learmonth Solar Observatory, Udaipur Solar Observatory, Instituto de
Astrof\'{\i}sica de Canarias, and Cerro Tololo Interamerican
Observatory. NSO/Kitt Peak data used here were produced cooperatively
by NSF/NOAO, NASA/GSFC, and NOAA/SEL; SOLIS data are
produced cooperatively by NSF/NSO and NASA/LWS. SOHO is a project of international cooperation between ESA and NASA. HMI data courtesy of NASA/SDO and the HMI science team. 
RH thanks the National Solar Observatory for computing support. GRD, YPE, and RH acknowledge the support of the UK Science and
Technology Facilities Council (STFC). 

\vspace{5mm}
\facilities{SOHO(MDI), SDO(HMI), NSO(GONG,KPVT,SOLIS)}

\end{document}